\documentclass[manuscript, table]{acmart}
\AtBeginDocument{%
  \providecommand\BibTeX{{%
    \normalfont B\kern-0.5em{\scshape i\kern-0.25em b}\kern-0.8em\TeX}}}

\usepackage{todonotes}
\presetkeys%
    {todonotes}%
    {inline,backgroundcolor=yellow}{}
\usepackage{verbatim}

\copyrightyear{2023} 
\acmYear{2023} 
\setcopyright{rightsretained} 
\acmConference[RecSys '23]{Seventeenth ACM Conference on Recommender Systems}{September 18--22, 2023}{Singapore, Singapore}
\acmBooktitle{Seventeenth ACM Conference on Recommender Systems (RecSys '23), September 18--22, 2023, Singapore, Singapore}
\acmDOI{10.1145/3604915.3610639}
\acmISBN{979-8-4007-0241-9/23/09}

\usepackage{moreverb}
\usepackage{xspace}
\usepackage{multirow}
\usepackage{xcolor}

\newcommand{\bert}{\textsc{\small{BERT}}\xspace}
\newcommand{\bertrec}{\textsc{\small{BERT4Rec}}\xspace}
\newcommand{\sasrec}{\textsc{\small{SASRec}}\xspace}
\newcommand{\grurec}{\textsc{\small{GRU4Rec}}\xspace}
\newcommand{\sknn}{\textsc{\small{SKNN}}\xspace}

\newcommand{\vsknn}{\textsc{\small{V\_SKNN}}\xspace}

\newcommand{\dhr}{Delivery Hero\xspace}

\newcommand{\embeddingModel}{\textsc{\small{LLMSeqSim}}}
\newcommand{\promptModel}{\textsc{\small{LLMSeqPrompt}}}
\newcommand{\bertmodel}{\textsc{\small{LLM2BERT4Rec}}}

\begin{document}

\title{Leveraging Large Language Models for Sequential Recommendation}

\author{Jesse Harte}
\affiliation{%
  \institution{Delivery Hero Research}
  \city{Berlin}
  \country{Germany}
}
\affiliation{%
  \institution{Delft University of Technology}
  \city{Delft}
  \country{The Netherlands}
}

\author{Wouter Zorgdrager}
\affiliation{%
  \institution{Delivery Hero Research}
  \city{Berlin}
  \country{Germany}
}

\author{Panos Louridas}
\affiliation{%
  \institution{Athens University of Economics \& Business}
  \city{Athens}
  \country{Greece}
}

\author{Asterios Katsifodimos}
\affiliation{%
  \institution{Delft University of Technology}
  \city{Delft}
  \country{The Netherlands}
}

\author{Dietmar Jannach}
\affiliation{%
  \institution{University of Klagenfurt}
  \city{Klagenfurt}
  \country{Austria}
}

\author{Marios Fragkoulis}
\affiliation{%
  \institution{Delivery Hero Research}
  \city{Berlin}
  \country{Germany}
}

\renewcommand{\shortauthors}{Harte et al.}

\begin{abstract}
Sequential recommendation problems have received increasing attention in research during the past few years, leading to the inception of 
a large variety of algorithmic approaches.
In this work, we
explore how large language models (LLMs), which are nowadays introducing
disruptive effects in many AI-based applications, can be used to build or improve sequential recommendation approaches. Specifically, we devise and evaluate three
approaches
to leverage the power of LLMs in different ways. Our results from experiments on two datasets show that initializing the state-of-the-art sequential recommendation model BERT4Rec with embeddings obtained from an LLM
improves NDCG by 15-20\% compared to the vanilla BERT4Rec model. Furthermore, we find that a simple approach that leverages LLM embeddings for producing recommendations, can provide competitive performance by highlighting semantically related items.
We publicly share the code and data of our experiments to ensure reproducibility.\footnote{\url{https://github.com/dh-r/LLM-Sequential-Recommendation}}
\end{abstract}




\maketitle

\section{Introduction}
\label{sec:introduction}
Sequential recommendation problems have received increased interest recently \cite{QuadranaetalCSUR2018,Wang2021survey}. In contrast to the traditional, sequence-agnostic matrix-completion setup \cite{Resnick:1994:GOA:192844.192905}, the problem in sequential recommendation is to predict the next user interest or action, given a sequence of past user interactions. Practical applications of sequential recommendation include next-purchase prediction, next-track music recommendation, or next Point-of-Interest suggestions for tourism. Due to their high practical relevance, a multitude of algorithmic approaches have been proposed in the past few years \cite{Hidas2016session,Ludewig2018,Kang2018selfattentive,Sun2019BERT4Rec}, including approaches that utilize side information about the items, such as an item's category
\cite{xu2022category,lai2022attribute}.

From a technical perspective, the sequential recommendation problem shares similarities with the next word prediction problem~\cite{radford2018improving, Devlin2019BERT}.
Under this light, we can observe a parallel between research in Natural Language Processing (NLP) and sequential recommendation, where novel recommendation models are inspired by NLP models~\cite{DeSouza2021Transformers4Rec}.
\grurec~\cite{Hidas2016session} adopted the Gated Recurrent Unit (GRU) mechanism from~\cite{cho2014learning}, \sasrec~\cite{Kang2018selfattentive} used the transformer architecture from~\cite{Vaswani2017attention}, and \bertrec~\cite{Sun2019BERT4Rec} adopted \bert~\cite{Devlin2019BERT}.
The influence of NLP research to sequential recommendation models extends naturally to Large Language Models (LLMs). LLMs, in particular ones based on Generative Pretrained Transformers~\cite{radford2018improving}, are
exhibiting disruptive effects in various AI-based applications with their semantically rich and meaningful responses.

However, limited research  exists so far
on leveraging the inherent semantic information of LLMs, which the abovementioned approaches lack, for sequential recommendation problems. A number of recent works in fact started to explore the potential of relying on LLMs for recommendation tasks; see \cite{lin2023recommender,Wu2023survey} for recent surveys. Here, we extend this line of research for sequential recommendation problems, providing the following contributions and insights.

\begin{itemize}
    \item We devise three orthogonal methods of leveraging LLMs for sequential recommendation. In our first approach (\embeddingModel), we retrieve a semantically-rich embedding from an existing LLM (from OpenAI) for each item in a session. We then compute an aggregate session embedding to recommend catalog products with a similar embedding. In the second approach (\promptModel), we fine-tune an LLM with dataset-specific information in the form of prompt-completion pairs and ask the model to produce next item recommendations for test prompts. Finally, our third approach (\bertmodel) consists of initializing existing sequential models with item embeddings obtained from an LLM.
    \item Experiments on two datasets, including a real-world dataset from \dhr, reveal that
    initializing a sequential model with LLM embeddings is particularly effective: applying it to the state-of-the-art model \bertrec improves accuracy in terms of NDCG by 15-20\%, making it the best-performing model in our experiments.
    \item Finally, we find that in certain applications simply using LLM embeddings to find suitable items for a given session (\embeddingModel) can lead to state-of-the-art performance.
\end{itemize}

\section{Background \& Related Work}
\label{sec:background}

The recent developments in LLMs have taken the world by surprise.
Models like OpenAI GPT~\cite{Brown2020Language}, Google BERT~\cite{Devlin2019BERT}, and Facebook LLaMA~\cite{touvron2023llama}, which employ deep transformer architectures, demonstrate how innovations in
NLP
can reshape mainstream online activities, such as search, shopping, and customer care.
Inevitably, research in recommender systems is significantly impacted by the developments in the area of LLMs as well.
According to recent surveys~\cite{lin2023recommender, Wu2023survey}, LLMs are mainly utilized for recommendation problems in two ways: by providing embeddings that can be used to initialize existing recommendation models~\cite{Wu2021Empowering, Zhang2021unbert, Liu21pretrained}, and by producing recommendations leveraging their inherent knowledge encoding~\cite{kang2023do, Hegselmann23tabllm, bao2023tallrec}.
LLMs as recommendation models can provide recommendations given \emph{a)} only a task specification (zero-shot), \emph{b)} a few examples given inline to the prompt of a task (few-shot), or \emph{c)} after fine-tuning the model's weights for a task given a set of training examples~\cite{Brown2020Language}.
This incremental training process deviates from typical recommendation models, which have to be trained from zero on domain data.
In fact, LLMs show early indications of 
adaptability to different recommendation domains with modest fine-tuning~\cite{hou2022towards, Hou2023learning}.
Finally, LLMs have been applied in various recommendation tasks, such as rating prediction~\cite{li2023text}, item generation~\cite{li2023gpt4rec}, and reranking~\cite{hou2023large} across domains~\cite{Wu2021Empowering, Liu21pretrained}.

In this work we explore the potential of using LLMs for sequential recommendation problems~\cite{jannach2020research}.
In short, in sequential recommendation problems, we consider as input a sequence of user interactions $S^u = (S^u_1, S^u_2, ..., S^u_n)$, for user $u$, where $n$ is the length of the sequence and $S^u_i$ are individual items. The aim is to predict the next interaction of the given sequence.
Besides the recent sequential recommendation models mentioned in the introduction~\cite{Hidas2016session, Kang2018selfattentive, Sun2019BERT4Rec}, in earlier works, the sequential recommendation problem has been modelled as a Markov Chain~\cite{Garcin2013personalized} or a Markov Decision Process~\cite{Shani2005MDP}.
Neighborhood-based approaches, such as SKNN~\cite{Jannach2017when}, have also been proposed.

Early research work regarding LLMs for sequential recommendation problems has shown mixed results
~\cite{Zhang2021language, Geng2022Recommendation, Ding2022zero, liu2023chatgpt, hou2023large}.
The very recent VQ-Rec model~\cite{Hou2023learning} employs a transformer architecture and applies a novel representation scheme to embeddings retrieved from \bert{} in order to adapt to new domains.
VQ-Rec outperforms a number of sequential recommendation models across datasets of different domains, and it has been shown that \sasrec with LLM embeddings is better than the original \sasrec method for half of the datasets representing different domains.
Finally, in an upcoming work~\cite{yuan2023recommender}, \sasrec with LLM embeddings is shown to improve over \sasrec.
The recent approaches presented in~\cite{Hou2023learning} and ~\cite{yuan2023recommender} differ from our work in particular in terms of the goals they pursue. VQ-Rec~\cite{Hou2023learning} targets cross-domain recommendations with a novel item representation scheme, while~\cite{yuan2023recommender} evaluates whether recommendation models leveraging different modalities perform better than existing recommendation models that rely on item identifiers.

The work presented in this paper complements these recent lines of research and proposes and evaluates three alternative ways of leveraging LLMs for sequential recommendation. Differently from earlier approaches, our work shows that initializing an existing sequential model with LLM-based embeddings is highly effective and helps to outperform existing state-of-the-art models. In addition, we find that retrieving relevant items solely based on LLM embedding similarity can lead to compelling recommendations depending on the dataset.

\section{Three LLM-based Approaches for Sequential Recommendations}
\label{sec:technical-approaches}
In this section, we describe the three technical approaches sketched in Section \ref{sec:introduction}.

\subsection{\embeddingModel: Recommending Semantically Related Items via LLM Embeddings}
\label{sub:LLM-embeddings}
With this first approach, our goal is to explore if recommendations can benefit from a holistic notion of similarity provided by LLMs.
To achieve this, we leverage \emph{LLM embeddings} to produce recommendations in three steps. First, we query the \texttt{text-embedding-ada-002}\footnote{\url{https://platform.openai.com/docs/guides/embeddings/second-generation-models}} OpenAI embedding model with the names of the products in the item catalog and retrieve their embeddings.
Second, we compute a session embedding for each session in our test set by combining the embeddings of the individual products in the session. Here, we try different combination strategies: \emph{a)} the average of the product embeddings, \emph{b)} a weighted average using linear and exponential decay functions depending on the position of the item in the session, and \emph{c)} only the embedding of the last product.\footnote{We also tried to create an aggregated session embedding by concatenating the plain product names and then querying the Open AI embeddings API. This however led to worse results.}
Third, we compare the session embedding to the embeddings of the items in the product catalog using cosine, Euclidean, and dot product similarity.\footnote{The choice of the similarity measure did not significantly impact the results.}
Finally, we recommend the top-\emph{k} products from the catalog with the highest embedding similarity to the session embedding.

\subsection{\promptModel: Prompt-based Recommendations by a Fine-Tuned LLM}
\label{sub:finetuned-LLM}

In this approach, we inject domain knowledge to the collective information that a base LLM incorporates, with the goal of increasing the quality of the recommendations by an LLM that is
given information about an ongoing session in the form of a prompt.
To this end, we fine-tune an OpenAI \texttt{ada}
model on training samples consisting of a prompt (the input) and a completion (the intended output). In our case, the prompt is a session, which contains a list of product names except for the last product, and the completion is the name of the last product in the same session, see Figure~\ref{fig:prompt-and-completion-example}.

To optimize
performance, we fine-tune the model until the validation loss converges.
After training, we provide the prompts of the sessions in the test set to the fine-tuned model to obtain recommendations.
We note that we make no strong assumption regarding the order of the returned recommendations.
Therefore, we use
the tendency of the model to provide duplicate recommendations as a proxy of its confidence and rank the recommendations by frequency of appearance. Then, to create
a full slate of unique recommendations, we retrieve the embedding of each duplicate product using the OpenAI embeddings API and take the catalog's product that is closest in terms of embedding similarity
using the dot product measure.
Finally,
we note that the fine-tuned LLM, being a generative model, may also return hallucinated products, which we map
to catalog products using the same method as for duplicate products.

\begin{figure}[h!t]
\begin{footnotesize}
\begin{boxedverbatim}
{"prompt":     "1. Burt's Bees Rosewater Toner 8oz\n
                2. Philosophy Lip Shine, Mimosa, 0.5 Ounce\n
                3. LaLicious Sugar Souffle Body Scrub 16 fl oz.\n
                4. Marc Jacobs Daisy Eau So Fresh Eau de Toilette Spray-125ml/4.25 oz.\n \n\n###\n\n",
 "completion": " Bcbg Max Azria Eau De Parfum Spray for Women, 3.4 Ounce ###"}
\end{boxedverbatim}
\end{footnotesize}
\caption{Example prompt and completion for fine-tuning from the Beauty dataset}
\label{fig:prompt-and-completion-example}
\end{figure}

\subsection{\bertmodel: Recommending with an LLM-enhanced Sequential Model}
\label{sub:bert-LLM-embeddings}
In our third approach, our goal is to leverage the semantically-rich item representations provided by an LLM to enhance an existing sequential recommendation model. Specifically, in our work we focus on \bertrec~\cite{Sun2019BERT4Rec}, a state-of-the-art transformer-based model,
which employs the transformer architecture~\cite{Vaswani2017attention} of \bert{}~\cite{Devlin2019BERT}.

\bert's transformer architecture consists of an embedding layer, a stack of encoder layers, and a projection head. Furthermore, \bert features a masked language model training protocol,
which involves masking items at random positions and letting the model predict their true identity. Initially, the embedding layer embeds an input sequence of (potentially masked) item IDs into a sequence of embeddings using both the item ID and the item position. Then the transformer encoder layers process the embedding sequence using a multi-head attention module and a feed-forward network shared across all positions.
Finally, the projection head projects the embeddings at each masked position to a probability distribution in order to obtain the true identity of the masked item. The projection head reuses the item embeddings of the embedding layer to reduce the model's size and to avoid
overfitting.

To allow \bertrec to leverage the rich information encoded in LLMs, 
we initialize \bertrec's item embeddings using the LLM embeddings described in Section~\ref{sub:LLM-embeddings}.
In order to align the embedding dimension of the LLM embeddings (1536) with the configured dimension of \bertrec's embedding layer (e.g., 64),
we employ Principal Components Analysis (PCA) to get 64 principal components of the LLM embeddings, which we then use to initialize the item embeddings of \bertrec's embedding layer.
Finally, we train the enhanced model the same way as our baseline \bertrec model.

\section{Experimental Evaluation}
\label{sec:experiments}
In this section, we describe our experimental setup (Section~\ref{subsec:experimental-setup}) and the results of our empirical evaluation (Section~\ref{subsec:results}).
\subsection{Experimental setup}
\label{subsec:experimental-setup}

\paragraph{Datasets and Data Splitting}
We use the public Amazon Beauty~\cite{He2016Ups} dataset
and a novel, real-world e-commerce dataset from \dhr\footnote{\url{https://www.deliveryhero.com}} for our experiments.
The Beauty dataset contains product reviews and ratings from Amazon. In line with prior research~\cite{Anelli2022Top}, we pre-processed the dataset to include at least five interactions per user and item (p-core = 5). 
The \dhr dataset contains anonymous QCommerce sessions for dark store and local shop orders. To better simulate a real-world setting, we did not preprocess this dataset, except that we removed sessions with only one interaction from the test set.
QCommerce is a segment of e-Commerce focusing on fast delivery times on the last mile. Dataset statistics are given in Table~\ref{tab:dataset-statistics}.
To create a train and test set in a sound way, we first split a dataset containing sessions temporally such that all test sessions succeed train sessions in time.
Then in the test set, we adopt the leave-one-out approach followed by~\cite{Kang2018selfattentive, Sun2019BERT4Rec} where all but the last interaction of each session represent the prompt, while the last interaction serves as the ground truth.

\begin{table}[t]
\begin{tabular}{rrrrrr}
\hline
\textbf{Dataset} & \textbf{\# sessions} & \textbf{\# items} & \textbf{\# interactions} & \textbf{Avg. length} & \textbf{Density} \\
\hline
Beauty 5-core &
22,363 &
12,101 &
198,502 &
8.9 &
0.073\%
\\
\hline
\dhr &
258,710 &
38,246 &
1,474,658 &
5.7 &
0.015\%
\\
\hline
\end{tabular}
\caption{Dataset statistics}
\label{tab:dataset-statistics}
\end{table}

\begin{figure}[t]
    \centering
    \begin{minipage}{0.4\textwidth}
        \centering
        \includegraphics[width=0.95\textwidth]{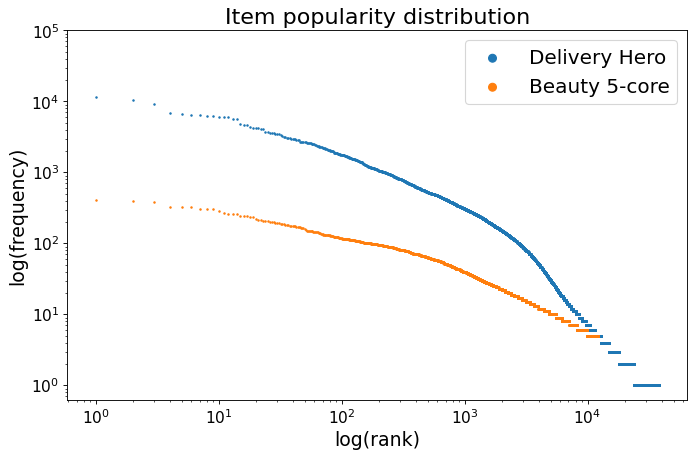} 
    \end{minipage}
    \begin{minipage}{0.4\textwidth}
        \centering
        \includegraphics[width=0.95\textwidth]{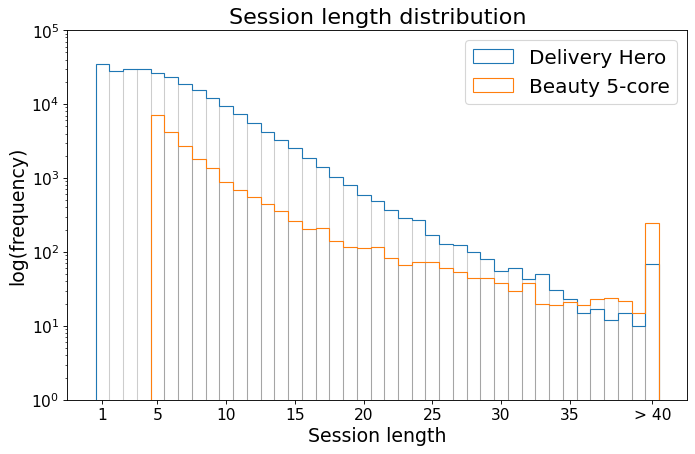} 
    \end{minipage}
    \caption{Distribution of items ranked by popularity (left) and histogram of session length (right) for the datasets}
    \label{fig:dataset-characteristics}
\end{figure}

\paragraph{Metrics}

We use the standard ranking accuracy metrics NDCG, MRR, and HitRate at the usual cut-off lengths of 10 and 20.
Furthermore, we consider the following \emph{beyond-accuracy} metrics to obtain a more comprehensive picture of the performance of the different algorithms: catalog coverage, serendipity, and novelty.
\emph{Catalog coverage} represents the fraction of catalog items that appeared in at least one top-n recommendation list of the users in the test set \cite{JannachLercheEtAl2015}. \emph{Serendipity} measures the average number of correct 
recommendations for each user that are not recommended by a popularity baseline~\cite{Ge2010beyond}.
\emph{Novelty} computes the negative log of the relative item popularity, or self-information~\cite{Zhou2010solving}.

\paragraph{Models}
We include both session-based algorithms of different families,  \grurec~\cite{Hidas2016session}, and \sknn{}~\cite{Jannach2017when}, as well as two state-of-the-art sequential models,  \bertrec~\cite{Sun2019BERT4Rec} and \sasrec~\cite{Kang2018selfattentive}.
We tested all variants of the \sknn nearest-neighbor method proposed in \cite{Ludewig2018} and report the results in the online material.
In addition,
we include the three LLM-based approaches proposed in Section~\ref{sec:technical-approaches}.
Finally, we include a popularity-based baseline (MostPopular) in the experiments. 
\paragraph{Hyperparameter Tuning} We systematically tuned all models (except the \embeddingModel{} and the \promptModel{}) on three validation folds with the Tree Parzen Estimator (TPE) sampler~\cite{Bergstra2011Algorithms}, and used the average NDCG@20 across the folds as the optimization goal.
For \promptModel{}, we applied manual hyperparameter search.
The examined hyperparameter ranges and optimal values for each dataset are reported in the online material.

\subsection{Results and Discussion}
\label{subsec:results}
Table~\ref{tab:results-beauty} and Table~\ref{tab:results-DH} show the results obtained for the Amazon Beauty and the \dhr dataset on the hidden test set, respectively. We report the best results of 5 runs.
The table is sorted according to NDCG@20.

\paragraph{Accuracy Results.}

The highest values in terms of NDCG@20 are obtained by \textbf{\bertmodel{}} for both datasets. In both cases, the gains obtained by using LLM-based item embeddings are substantial,
demonstrating the benefits of relying on semantically-rich embeddings in this sequential model. The NDCG value increased by more than 20\% for Beauty 
and over 15\% on the \dhr dataset.\footnote{We also examined the value of LLM embeddings for the \sasrec model, where we observed marked increases in the NDCG, but not to the extent that it outperformed \bertmodel{}. We report these additional results in the online material.} 
To confirm that the semantics of the LLM embeddings is the driver of performance, we ran an experiment in which we permuted the item embeddings such that the embedding of each item is initialized to the principal components of the LLM embedding of another product from the catalogue. The experiment maintains the statistical properties of the embeddings, but deprives the item embeddings of the semantics of the LLM embeddings. The resulting model exhibited worse performance than the baseline \bertrec model with randomly-initialized item embeddings clearly showing that the performance improvement cannot be credited to the statistical properties of the embeddings.

The relative performance of \textbf{\embeddingModel{}}, again considering NDCG values, varies across the two datasets. On the Beauty dataset, the model is highly competitive, with NDCG@20 values only being slightly lower than \bertmodel. At shorter list lengths, i.e., at NDCG@10, the \embeddingModel{} model even leads to the best performance for this dataset. Notably, the embedding combination strategy that led to the best results considered only the last item of the session (see Section~\ref{sub:LLM-embeddings}). For the \dhr dataset, in contrast, the picture is very different, and \embeddingModel{} leads to quite poor performance, only outperforming the popularity-based baseline.
We hypothesize that this phenomenon is a result of the
quite different characteristics of the two datasets. For example, in Figure~\ref{fig:dataset-characteristics}, we observe that many items in the real-world \dhr dataset occur very infrequently.
This may limit the capacity of \embeddingModel{} to find
similar items, given also the substantially broader item catalog
in the \dhr dataset. Furthermore, a manual inspection of a sample of test prompts, recommendations, and ground truths of the two datasets indicates that users in the Beauty dataset frequently rate items of a certain brand. Since brand names are part of the product names that are input
to the LLM, recommending similar items may turn out to be particularly effective.

\begin{table}[t]
\resizebox{\textwidth}{!}{
\begin{tabular}{lrrrrrr|rrrrrr}
\hline
\multicolumn{1}{c}{} &
\multicolumn{6}{c}{\textbf{Top@10}} &
\multicolumn{6}{c}{\textbf{Top@20}}                                                                   \\
\multicolumn{1}{c}{\multirow{-2}{*}{\textbf{Model}}} &
\textbf{nDCG} &
\textbf{HR} &
\textbf{MRR} &
\textbf{CatCov} &
\textbf{Seren} &
\textbf{Novel} &
\underline{\textbf{nDCG}} &
\textbf{HR} &
\textbf{MRR} &
\textbf{CatCov} &
\textbf{Seren} &
\textbf{Novel}
\\
\hline
\rowcolor{gray!20}
\bertmodel &
    0.041 &
    \textbf{0.076} &
    0.030 &
    0.180 &
    \textbf{0.072} &
    11.688 &
    \textbf{0.051} &
    \textbf{0.118} &
    0.033 &
    0.260 &
    \textbf{0.110} &
    11.888
\\
\embeddingModel &
    \textbf{0.044} &
    0.063 &
    \textbf{0.038} &
    \textbf{0.763} &
    0.063 &
    \textbf{13.819} &
    0.048 &
    0.079 &
    \textbf{0.039} &
    \textbf{0.889} &
    0.079 &
    \textbf{13.858}
\\
\rowcolor{gray!20}
\vsknn &
    0.041 &
    0.071 &
    0.033 &
    0.673 &
    0.069 &
    12.241 &
    0.047 &
    0.095 &
    0.034 &
    0.889 &
    0.091 &
    12.492
\\
\bertrec &
    0.034 &
    0.067 &
    0.024 &
    0.231 &
    0.064 &
    12.293 &
    0.043 &
    0.103 &
    0.027 &
    0.312 &
    0.098 &
    12.423
\\
\rowcolor{gray!20}
\grurec &
    0.027 &
    0.051 &
    0.020 &
    0.145 &
    0.047 &
    11.409 &
    0.035 &
    0.082 &
    0.022 &
    0.214 &
    0.074 &
    11.597
\\
\sasrec &
    0.026 &
    0.051 &
    0.019 &
    0.121 &
    0.048 &
    11.485 &
    0.033 &
    0.080 &
    0.021 &
    0.182 &
    0.073 &
    11.678
\\
\rowcolor{gray!20}
\promptModel &
    0.025 &
    0.045 &
    0.019 &
    0.500 &
    0.044 &
    13.001 &
    0.030 &
    0.064 &
    0.020 &
    0.688 &
    0.063 &
    13.361
\\
MostPopular &
    0.005 &
    0.010 &
    0.003 &
    0.001 &
    0.001 &
    9.187 &
    0.006 &
    0.018 &
    0.003 &
    0.002 &
    0.001 &
    9.408
\\
\hline
\end{tabular}
}
\caption{Evaluation results for the Amazon Beauty dataset}
\label{tab:results-beauty}
\end{table}

\begin{table}[t]
\resizebox{\textwidth}{!}{
\begin{tabular}{lrrrrrr|rrrrrr}
\hline
\multicolumn{1}{c}{} &
\multicolumn{6}{c}{\textbf{Top@10}} &
\multicolumn{6}{c}{\textbf{Top@20}}                                                                   \\
\multicolumn{1}{c}{\multirow{-2}{*}{\textbf{Model}}} &
\textbf{nDCG} &
\textbf{HR} &
\textbf{MRR} &
\textbf{CatCov} &
\textbf{Seren} &
\textbf{Novel} &
\underline{\textbf{nDCG}} &
\textbf{HR} &
\textbf{MRR} &
\textbf{CatCov} &
\textbf{Seren} &
\textbf{Novel}
\\
\hline
\rowcolor{gray!20}
\bertmodel &
    \textbf{0.102} &
    \textbf{0.179} &
    \textbf{0.078} &
    0.245 &
    \textbf{0.151} &
    10.864 &
    \textbf{0.120} &
    \textbf{0.252} &
    \textbf{0.083} &
    0.311 &
    \textbf{0.198} &
    11.050
\\
\bertrec &
    0.088 &
0.157 &
0.067 &
0.325 &
0.128 &
10.821 &
0.104 &
0.221 &
0.071 &
0.429 &
0.165 &
11.032
\\
\rowcolor{gray!20}
\grurec &
    0.085 &
    0.153 &
    0.064 &
    0.127 &
    0.124 &
    10.570 &
    0.101 &
    0.218 &
    0.068 &
    0.172 &
    0.161 &
    10.823
\\
\sasrec &
    0.084 &
0.149 &
0.065 &
0.170 &
0.120 &
10.674 &
0.100 &
0.212 &
0.069 &
0.229 &
0.156 &
10.913
\\
\rowcolor{gray!20}
\vsknn &
    0.087 &
0.148 &
0.068 &
0.381 &
0.120 &
10.444 &
0.100 &
0.200 &
0.072 &
0.452 &
0.146 &
10.602
\\
\promptModel &
    0.063 &
    0.116 &
    0.047 &
    0.400 &
    0.107 &
    12.048 &
    0.070 &
    0.144 &
    0.049 &
    0.611 &
    0.123 &
    13.788
\\
\rowcolor{gray!20}
\embeddingModel &
    0.039 &
    0.069 &
    0.029 &
    \textbf{0.633} &
    0.069 &
    \textbf{16.315} &
    0.046 &
    0.096 &
    0.031 &
    \textbf{0.763} &
    0.093 &
    \textbf{16.536}
\\
MostPopular &
    0.024 &
    0.049 &
    0.017 &
    0.000 &
    0.000 &
    7.518 &
    0.032 &
    0.079 &
    0.019 &
    0.001 &
    0.000 &
    7.836
\\
\hline
\end{tabular}
}
\caption{Evaluation results for the \dhr dataset}
\label{tab:results-DH}
\end{table}

Looking at the other accuracy metrics (\textbf{Hit Rate} and \textbf{MRR}), we find that these are generally highly correlated with the NDCG results. A notable exception are the MRR values of the \embeddingModel{} model and the \vsknn{} approach
on the Beauty dataset. While these two approaches lead to slightly inferior results at NDCG@20 and in particular also for HR@20, they are superior in terms of MRR. This means that these methods place the hidden target item higher up in the recommendation list in case the target item is included in the top 20.
Similar observations regarding the good performance of some methods in terms of MRR on specific datasets were previously reported also in \cite{Ludewig2018}.

Interestingly, as also reported in \cite{Ludewig2018,latifi2022sequential}, \textbf{nearest-neighbor} approaches can be quite competitive depending on the dataset.
On Beauty, 
\vsknn{} outperforms all of the more sophisticated neural models (\bertrec, \grurec, \sasrec) in all accuracy metrics except Hit Rate@20.
On the \dhr dataset, in contrast, 
the neural models perform better in all accuracy metrics except MRR and NDCG@10.
Further inspection (see online material)
showed that \sknn's performance drops as the length of sessions increases, while the performance of the other models remains stable.

The performance of the \textbf{\promptModel{}} model
again depends on the dataset. On the Beauty dataset, it leads to accuracy  values that are often only slightly lower than \sasrec{}, which
is
typically considered a strong state-of-the-art baseline. On the \dhr dataset, in contrast, the drop in performance compared to the other models is substantial.
Still, \promptModel{} leads to accuracy values that are markedly
higher than the popularity baseline.
Given its versatility, ease of configuration and promising performance, \promptModel{} merits further research.

\paragraph{Beyond-Accuracy Results.} We make the following observations for \textbf{coverage}, \textbf{serendipity} and \textbf{novelty}. The \embeddingModel{} model consistently leads to the best coverage and novelty.
This is not too surprising, given the nature of the approach, which is solely based on embeddings similarities. Unlike other methods that use collaborative signals, i.e., past user-item interactions, the general popularity of an item in terms of the amount of observed past interactions does not play role in \embeddingModel{}, neither directly nor implicitly. Thus, the model has no
tendency to concentrate the recommendations on a certain subset of (popular) items. We recall that the used novelty measure is based on the popularity of the items in the recommendations.
The serendipity results are largely aligned with the accuracy measures across the datasets. This generally confirms the value of personalizing the recommendations to individual user preferences, compared to recommending mostly popular items to everyone. We iterate that our serendipity measure counts the fraction of correctly recommended items that would not be recommended by a popularity-based approach.

\section{Conclusions}

In this work, we devised and evaluated three approaches that leverage LLMs for sequential recommendation problems. A systematic empirical evaluation revealed that \bertrec initialized with LLM embeddings achieves the best performance for two datasets, and that the LLM-based initialization leads to a substantial improvement in accuracy.
In our future work, we plan to investigate if our findings generalize to different domains, using alternative datasets with diverse characteristics. Furthermore, we will explore if using other LLMs, e.g., ones with different architectures and training corpora, will lead to similar performance gains, including a hybrid of \bertmodel{} with \embeddingModel{} towards combining their accuracy and beyond-accuracy performance. Finally, it is open so far if passing other types of information besides product names, e.g., category information, to an LLM can help to further improve
the performance of the models.


\bibliographystyle{ACM-Reference-Format}
\bibliography{llm-reco}

\end{document}